\documentclass[12pt, a4paper]{article}
\baselineskip=\normalbaselineskip

\usepackage{mathrsfs,amsbsy,amssymb,latexsym,amsfonts,amsmath,amsthm}
\usepackage{graphicx,color}
\usepackage[nosort]{cite}
\usepackage{algpseudocode}
\usepackage{algorithm}
\usepackage{bm}
\usepackage{algorithm}
\usepackage{algpseudocode}

\newcommand{\bbC}{{\mathbb{C}}}
\newcommand{\bbR}{{\mathbb{R}}}
\newcommand{\bbZ}{{\mathbb{Z}}}
\newcommand{\calF}{{\mathcal{F}}}
\newcommand{\calJ}{{\mathcal{J}}}
\newcommand{\calK}{{\mathcal{K}}}

\newcommand{\calO}{{\mathcal{O}}}
\newcommand{\calR}{{\mathcal{R}}}
\newcommand{\ReS}{{\rm Re}\,S}
\newcommand{\ImS}{{\rm Im}\,S}

\newcommand{\Nconf}{{N_{\rm conf}}\,}
\newcommand{\circdot}[1]{{\overset{\circ}{#1}}}

\makeatletter
\catcode`\@=11
\@addtoreset{equation}{section}

\def\@seccntformat#1{\csname the#1\endcsname.~~}
\makeatother

\makeatletter
\catcode`\@=11
\@addtoreset{equation}{section}

\def\@seccntformat#1{\csname the#1\endcsname.~~}
\makeatother

\begin{document}

\begin{titlepage} 

\renewcommand{\thefootnote}{\fnsymbol{footnote}}
\begin{flushright}
  KUNS-2987
\end{flushright}
\vspace*{1.0cm}

\begin{center}
{\Large \bf
Simplified algorithm for the Worldvolume HMC\\
and the Generalized-thimble HMC
}
\vspace{1.0cm}

\centerline{
{Masafumi Fukuma}%
\footnote{
  E-mail address: fukuma@gauge.scphys.kyoto-u.ac.jp
}
}

\vskip 0.5cm
  {\it Department of Physics, Kyoto University,
  Kyoto 606-8502, Japan}

\end{center}

\begin{abstract}

The Worldvolume Hybrid Monte Carlo method (WV-HMC method) [arXiv:2012.08468] 
is a reliable and versatile algorithm towards solving the sign problem. 
Similarly to the tempered Lefschetz thimble method, 
this method removes the ergodicity problem 
inherent in algorithms based on Lefschetz thimbles. 
In addition to this advantage, 
the WV-HMC method significantly reduces the computational cost 
because one needs not compute the Jacobian of deformation 
in generating configurations. 
A crucial step in this method is the RATTLE algorithm, 
where the Newton method is used at each molecular dynamics step 
to project a transported configuration onto a submanifold (worldvolume) 
in the complex space. 
In this paper, 
we simplify the RATTLE algorithm 
by employing a simplified Newton method (the fixed-point method) 
along with iterative solvers for orthogonal decompositions of vectors, 
and show that this algorithm further reduces the computational cost. 
We also apply this algorithm 
to the HMC algorithm for the generalized thimble method (GT-HMC method). 
We perform a numerical test 
for the convergence of the simplified RATTLE algorithm, 
and show that the convergence depends on the system size only weakly. 
The application of this simplified algorithm to various models 
will be reported in subsequent papers.

\end{abstract}
\end{titlepage}

\pagestyle{empty}
\pagestyle{plain}

\tableofcontents
\setcounter{footnote}{0}

\section{Introduction}
\label{sec:intro}

The numerical sign problem has been a major obstacle to 
first-principles calculations in various important physical systems. 
Typical examples include 
finite-density QCD \cite{Aarts:2015tyj}, 
Quantum Monte Carlo calculations for strongly correlated electron systems 
and frustrated spin systems \cite{Pollet:2012}, 
and the real-time dynamics of quantum many-body systems. 

The sign problem has a long history, 
and specific algorithms have been developed so far 
for each system with the sign problem. 
However, in the last two decades 
there has been a movement to develop more versatile methods 
for solving the sign problem, 
and various algorithms have been proposed.
One of these is a class of algorithms based on the complex Langevin equation 
\cite{Parisi:1983cs,Klauder:1983sp,
Aarts:2008wh,Aarts:2009uq,Aarts:2011ax,Nagata:2016vkn}. 
Another is based on Lefschetz thimbles 
\cite{Witten:2010cx,Cristoforetti:2012su,Cristoforetti:2013wha,
Fujii:2013sra,Fujii:2015bua,Fujii:2015vha,Alexandru:2015xva,Alexandru:2015sua,
Fukuma:2017fjq,Alexandru:2017oyw,Alexandru:2017lqr,Fukuma:2019wbv,Alexandru:2019,
Fukuma:2019uot,Fukuma:2020fez,Fukuma:2021aoo} 
(the path optimization method 
\cite{Mori:2017pne,Mori:2017nwj,Alexandru:2018fqp,Bursa:2018ykf} 
may be included in this class). 
There has also been an intensive study of non-Monte Carlo techniques, 
such as the tensor network method 
\cite{Levin:2006jai,Xie:2009,Adachi:2019paf,Shimizu:2014uva,Akiyama:2020sfo}. 

As will be reviewed in Sect.~\ref{sec:sign_problem}, 
in the Lefschetz thimble method, 
one deforms the integration surface of the path integral 
into the complex space 
so that the sign problem is alleviated on the new integration surface. 
This algorithm has the advantage 
that correct convergence is guaranteed by the Picard-Lefschetz theory, 
and in principle it can be applied to any system 
so long as the system can be expressed with continuous variables.  
However, as will be also discussed in Sect.~\ref{sec:sign_problem}, 
naive Monte Carlo implementations lead to serious ergodicity problems 
\cite{Fujii:2015bua,Fujii:2015vha}. 

The tempered Lefschetz thimble method (TLT method) \cite{Fukuma:2017fjq} 
was introduced to solve the sign and ergodicity problems simultaneously, 
by implementing the tempering algorithm 
using the deformation parameter as the tempering parameter. 
The TLT method, however, has a drawback of high computational cost. 
In fact, one needs to compute the Jacobian of the deformation 
at every stochastic step along the direction of deformation, 
whose cost is $O(N^3)$ ($N$ is the number of degrees of freedom). 
To reduce the computational cost, 
the Worldvolume Hybrid Monte Carlo method (WV-HMC method) 
was invented in Ref.~\cite{Fukuma:2020fez}, 
where one considers molecular dynamics 
over a continuous accumulation of deformed surfaces (worldvolume). 
While retaining the advantages of the TLT method, 
the WV-HMC method significantly reduces the computational cost 
because it no longer needs the computation of the Jacobian 
in generating configurations. 

The main aim of this paper is to clarify and simplify the WV-HMC algorithm 
to a level 
at which it is accessible to a wider range of researchers. 
A crucial step in the WV-HMC method is the RATTLE algorithm 
\cite{Andersen:1983,Leimkuhler:1994}, 
which projects at each molecular dynamics step 
a transported configuration onto the worldvolume \cite{Fukuma:2020fez}. 
The projection requires solving constraint equations, 
which can be done with the Newton method. 
In this paper, 
we simplify the RATTLE algorithm 
by employing a simplified Newton method, 
and show that this algorithm further reduces the numerical cost. 
The introduction of a simplified Newton method to the Lefschetz thimble method 
was first made as the \emph{fixed-point method} 
in a seminal paper by the Komaba group \cite{Fujii:2013sra}, 
where the projection onto a single thimble is considered 
using the explicit form of the Jacobian.%
\footnote{
	The author thanks the referee 
	for the comment on the first version of the manuscript, 
	making him aware that  
	the simplified Newton method considered here is the same 
	as the fixed-point method of Ref.~\cite{Fujii:2013sra}.
}  
The fixed-point method saves us from solving extra linear equations 
that was necessary for the standard (non-simplified) Newton method 
\cite{Fukuma:2019uot,Fukuma:2020fez}. 
Furthermore, 
by combining the fixed-method 
with iterative solvers \cite{Alexandru:2017lqr} 
for orthogonal decompositions of vectors 
that use only integrations of flow equations, 
one no longer needs to compute the Jacobian explicitly 
unlike the original fixed-point method. 
We also apply this algorithm 
to the HMC algorithm for the generalized thimble method (GT-HMC method) 
\cite{Alexandru:2019,Fukuma:2019uot}. 
The application of the simplified WV-HMC algorithm to various models 
will be reported in subsequent papers \cite{fn1,fn2,mf}. 

This paper is organized as follows. 
In Sect.~\ref{sec:sign_problem}, 
we first define the sign problem, 
and briefly summarize various algorithms 
proposed so far based on Lefschetz thimbles. 
Section~\ref{sec:gt} deals with the simplification of the GT-HMC method. 
We show that the projection onto a deformed integration surface 
can be effectively performed by a simplified Newton method  
(fixed-point method) 
when combined with iterative solvers for orthogonal decompositions of vectors. 
This algorithm is extended to the WV-HMC method in Sect.~\ref{sec:wv}. 
In Sect.~\ref{sec:test}, 
we perform a numerical test 
for the convergence of the simplified RATTLE algorithm, 
and show that the convergence depends on the system size only weakly. 
Section~\ref{sec:conclusion} is devoted to conclusion 
and outlook for the application of the simplified algorithm to various models.

\section{The sign problem and various algorithms based on Lefschetz thimbles}
\label{sec:sign_problem}

Let $x=(x^i)\in\bbR^N$ be a dynamical variable 
with flat measure $dx\equiv dx^1\wedge\cdots\wedge dx^N$, 
and $S(x)$ and $\calO(x)$ the action and observables, respectively. 
Our aim is to estimate the expectation values of $\calO$ 
with respect to the Boltzmann weight 
$\rho(x)\equiv e^{-S(x)}/\int dx\,e^{-S(x)}$: 
\begin{align}
  \langle \calO \rangle \equiv 
  \int_{\bbR^N} dx\,\rho(x)\,\calO(x) 
  = \frac{\int_{\bbR^N} dx\,e^{-S(x)}\,\calO(x)}
  {\int_{\bbR^N} dx\,e^{-S(x)}}.
\label{vev}
\end{align}
When the action is complex-valued, 
one cannot regard $\rho(x)$ as a probability distribution, 
and a direct use of the Monte Carlo method is not possible. 
A standard way around is to reweight the integral 
with the real part of the action, $\ReS(x)$, 
and rewrite the integral 
as a ratio of reweighted averages:
\begin{align}
  \langle \calO \rangle
  = \frac{\langle e^{-i \ImS(x)}\,\calO(x) \rangle_{\rm rewt}}
         {\langle e^{-i \ImS(x)} \rangle_{\rm rewt}},
\label{rewt1}
\end{align}
where the reweighted average $\langle\cdots\rangle_{\rm rewt}$ is defined by
\begin{align}
  \langle g(x) \rangle_{\rm rewt} \equiv 
  \frac{\int dx\,e^{-\ReS(x)}\,g(x)}
       {\int dx\,e^{-\ReS(x)}}.
\label{rewt2}
\end{align}
The reweighted averages in Eq.~\eqref{rewt1} become highly oscillatory integrals 
with large degrees of freedom ($N\gg 1$), 
giving very small values of the form $e^{-O(N)}$. 
This should not be a problem 
if the reweighted averages can be estimated precisely, 
but in the Monte Carlo calculations 
they are accompanied by statistical errors of $O(1/\sqrt{\Nconf})$ 
for a sample of size $\Nconf$:
\begin{align}
  \langle \calO \rangle \approx
  \frac{e^{-O(N)} \pm O(1/\sqrt{\Nconf})}
       {e^{-O(N)} \pm O(1/\sqrt{\Nconf})},
\end{align}
and thus we need an exponentially large sample size, $\Nconf \gtrsim e^{O(N)}$, 
in order to make the statistical errors relatively smaller than the means. 
This is the sign problem we consider in this paper.

In the Lefschetz thimble method, 
the integration surface $\Sigma_0 = \bbR^N=\{x\}$ is continuously deformed 
into the complex space $\bbC^N=\{z=x+i\,y\}$
in such a way that the oscillatory behavior is alleviated 
on the deformed surface $\Sigma \in \bbC^N$. 
Throughout this paper, 
we assume that $e^{-S(z)}$ and $e^{-S(z)}\,\calO(z)$ are entire functions in $\bbC^N$ 
(which usually holds for systems of interest). 
Then, Cauchy's theorem ensures 
that the integrals do not change under deformations 
if the boundaries at ${\rm Re}\,z\to\pm\infty$ are kept fixed, 
and we have  
\begin{align}
  \langle \calO \rangle 
  &= 
  \frac{\int_\Sigma dz\, e^{-S(z)}\,\calO(z)}
       {\int_\Sigma dz\, e^{-S(z)}}.
\label{vev_Sigma}
\end{align}
We consider the deformation with the anti-holomorphic flow 
defined by the following flow equation:
\begin{align}
  \dot{z} = \overline{\partial S(z)} \quad 
  \bigl[ \partial S(z) = (\partial_i S(z))~(i=1,\ldots,N) \bigr].
\end{align}
This leads to the inequality 
$[S(z)]^\centerdot=(\partial S(z)/\partial z)\cdot \dot{z} = |\partial S(z)|^2\geq 0$, 
from which we know that $\ReS(z)$ always increases under the flow 
except at critical points $\zeta$
[where the gradient of the action vanishes, $\partial S(\zeta)$=0], 
while $\ImS(z)$ is kept constant. 
This flow sends the original integration surface $\Sigma_0=\bbR^N$ 
to a deformed surface $\Sigma_t$ at flow time $t$, 
which in the large flow time limit 
moves to a vicinity of a homological sum of Lefschetz thimbles:
\begin{align}
  \Sigma_t \to \sum_\sigma n_\sigma\,\calJ_\sigma
  \quad
  (n_\sigma\in\bbZ).
\end{align}
Here, $\sigma$ labels critical points, 
and $J_\sigma$ is the Lefschetz thimble associated with critical point $\zeta_\sigma$, 
which is defined as the union of orbits flowing out of $\zeta_\sigma$.%
\footnote{ 
  If we further introduce the anti-thimble $\calK_\sigma$ 
  as the union of orbits flowing \emph{into} $\zeta_\sigma$, 
  the coefficient $n_\sigma$ is expressed 
  as the intersection number between the original surface and the anti-thimble, 
  $n_\sigma = \langle \Sigma_0, \calK_\sigma \rangle$ 
  \cite{Witten:2010cx}.
} 
Since $\ImS(z)$ is constant on each Lefschetz thimble 
[$\ImS(z)=\ImS(z_\sigma)$ for $z\in\calJ_\sigma$], 
the oscillatory behavior of integrals at large flow times 
is expected to be much relaxed around each Lefschetz thimble. 

\begin{figure}[tb]
  \centering
  \includegraphics[width=90mm]{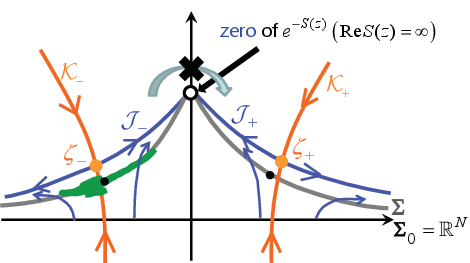}
  \caption{Ergodicity problem. 
    Configurations can hardly move from a vicinity of one thimble $\calJ_-$
    to that of another thimble $\calJ_+$.}
  \label{fig:ergodicity_problem}
\end{figure}

While the sign problem attributed to oscillatory integrals gets alleviated 
as we increase the flow time, 
there comes out another problem, the ergodicity problem. 
In fact, 
$\ReS(z)$ diverges at the boundaries of Lefschetz thimbles, 
which are zeros of the Boltzmann weight $\propto e^{-S(z)}$, 
and it is hard for configurations 
to move from a vicinity of one thimble 
to that of another thimble in stochastic processes 
(see Fig.~\ref{fig:ergodicity_problem}). 
Thus, we have a dilemma between the alleviation of the sign problem
and the emergence of the ergodicity problem. 

The \emph{generalized thimble} method \cite{Alexandru:2015sua} 
is an algorithm 
which makes a sampling on a deformed surface 
at such a flow time 
that is large enough to relax the oscillatory behavior 
and at the same time is small enough to avoid the ergodicity problem. 
However, a closer investigation \cite{Fukuma:2019wbv} 
shows that 
the oscillatory behaviors usually starts being relaxed 
only after the deformed surface reaches some of the zeros of $e^{-S(z)}$, 
so that one can hardly expect such an ideal flow time to be found.  
Nevertheless, this algorithm is still useful 
for grasping a flow time 
at which the sign problem starts being relaxed, 
by observing the average phase factors 
$\langle e^{-i \ImS(z)}\,dz/|dz| \rangle_{\Sigma_t}$ 
at various flow times. 
Configurations on (a connected component of) 
a deformed surface $\Sigma_t$ can be generated efficiently 
with the Hybrid Monte Carlo algorithm \cite{Alexandru:2019,Fukuma:2019uot}, 
which we refer in this paper 
to the \emph{Generalized-thimble Hybrid Monte Carlo} (GT-HMC) 
and review in the next section.%
\footnote{ 
  A HMC algorithm with RATTLE was first introduced 
  to the Lefschetz thimble method 
  in a monumental paper by the Komaba group \cite{Fujii:2013sra}, 
  where sampling is done directly on a single thimble.
} 

\begin{figure}[tb]
  \centering
  \includegraphics[width=90mm]{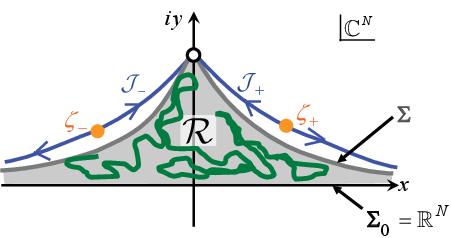}
  \caption{Worldvolume $\mathcal{R}$.}
  \label{fig:worldvolume}
\end{figure}

The \emph{tempered Lefschetz thimble} (TLT) method \cite{Fukuma:2017fjq} 
avoids the above dilemma 
by implementing the tempering algorithm 
with the flow time as the tempering parameter. 
This is the first algorithm 
that solves the sign and ergodicity problems simultaneously, 
but has a drawback of large computational cost 
[$O(N^3)$ for generating a configuration]. 
The \emph{Worldvolume Hybrid Monte Carlo} (WV-HMC) method \cite{Fukuma:2020fez}
was then introduced to reduce the computational cost of the TLT method 
while still retaining its advantages. 
This is based on the molecular dynamics 
on a continuous accumulation (worldvolume) of deformed surfaces 
(see Fig.~\ref{fig:worldvolume}). 
The TLT and WV-HMC methods have been successfully applied 
to $(0+1)$-dimensional Thirring model \cite{Fukuma:2017fjq}, 
the Hubbard model away from half filling \cite{Fukuma:2019wbv} 
and the Stephanov model \cite{Fukuma:2020fez} 
(although the system sizes are yet small).
%

\section{Generalized-thimble Hybrid Monte Carlo (GT-HMC)}
\label{sec:gt}

We explain the basics of GT-HMC \cite{Alexandru:2019,Fukuma:2019uot}%
\footnote{ 
  The GT-HMC algorithm is treated in Ref.~\cite{Fukuma:2019uot} 
  as part of the TLT method  
  and is combined with the parallel tempering algorithm 
  with the flow time as the tempering parameter. 
  The presentation below follows this reference.
} 
and propose its simplified algorithm. 
In the following, 
$\Sigma\equiv\Sigma_t$ is the deformed surface at flow time $t$, 
and $T_z \Sigma$ and $N_z\Sigma$ represent the tangent and the normal spaces 
at $z$ to $\Sigma$, respectively.

\subsection{Path-integral form for GT-HMC}
\label{sec:gt_pi}

We start from the expression [Eq.~\eqref{vev_Sigma}] 
\begin{align}
  \langle \calO \rangle 
  &= 
  \frac{\int_\Sigma dz\, e^{-S(z)}\,\calO(z)}
       {\int_\Sigma dz\, e^{-S(z)}}.
\label{vev_Sigma2}
\end{align}
With local coordinates%
\footnote{ 
  A canonical choice of $x$ is initial configurations of the flow, 
  but they can also be set to vectors in the tangent space 
  at a point on $\Sigma$ 
  as in Refs.~\cite{Fujii:2013sra} and \cite{Alexandru:2015sua} .
\label{fn:coords}
} 
$x=(x^a)$ $(a=1,\ldots,N)$ 
and the Jacobian $E^i_a \equiv \partial z^i/\partial x^a$, 
the holomorphic $N$-form $dz = dz^1\wedge\cdots\wedge dz^N$ 
is expressed as 
\begin{align}
  dz = \det E\,dx. 
\end{align}
We introduce the inner product
\begin{align}
  \langle u, v \rangle \equiv {\rm Re\,}u^\dagger v 
  = \sum_{i=1}^N {\rm Re\,} \overline{u^i} v^i
  \,\,\bigl( = \langle v, u \rangle \bigr)
\end{align}
for vectors $u=(u^i),\,v=(v^i)\in\bbC^N$. 
The induced metric 
$ds^2 = \gamma_{ab}\,dx^a dx^b \equiv |dz(x)|^2$ 
is then given by%
\footnote{ 
  We have used the identity 
  ${\rm Im}\,E_a^\dag E_b = 0$ \cite{Fujii:2013sra}, 
  which holds when the original configuration space $\Sigma_0$ is flat. 
} 
\begin{align}
  \gamma_{ab} = \langle E_a,E_b \rangle = E_a^\dag E_b,
\end{align}
which yields the invariant volume element on $\Sigma$, 
\begin{align}
  |dz| \equiv \sqrt{\gamma}\,dx = |\det E\,|\,dx. 
\end{align}
The expectation value \eqref{vev_Sigma2} is then expressed 
as a ratio of reweighted averages on $\Sigma$: 
\begin{align}
  \langle \calO \rangle 
  = \frac{\langle \calF(z)\,\calO(z) \rangle_\Sigma}
         {\langle \calF(z) \rangle_\Sigma},
\end{align}
where $\langle\cdots\rangle_\Sigma$ is defined by
\begin{align}
  \langle g(z) \rangle_\Sigma
  \equiv
  \frac{\int_\Sigma |dz|\,e^{-\ReS(z)}\,g(z)} 
       {\int_\Sigma |dz|\,e^{-\ReS(z)}}
\end{align}
and $\calF(z)$ is the associated reweighting factor:
\begin{align}
  \calF(z) \equiv \frac{dz}{|dz|}\,e^{-i\,\ImS(z)}
  = \frac{\det E}{|\det E\,|}\,e^{-i\,\ImS(z)}.
\end{align}

The reweighted average $\langle \cdots \rangle_\Sigma$  
can be written as a path integral over the phase space 
by rewriting the measure $|dz|=\sqrt{\gamma}\,dx$ to the form 
\begin{align}
  |dz| = \sqrt{\gamma}\,dx 
  \propto dx\,dp\,e^{-(1/2)\,\gamma^{ab}\,p_a p_b},
\end{align}
where $dx\,dp\equiv \prod_a \bigl( dx^a dp_a \bigr)$ 
is the volume element of the phase space 
and $(\gamma^{ab})\equiv (\gamma_{ab})^{-1}$. 
We thus obtain the phase-space path integral in the parameter-space representation: 
\begin{align}
  \langle g(z) \rangle_\Sigma 
  = \frac{\int dx\,dp\,e^{-(1/2)\,\gamma^{ab}\,p_a p_b - \ReS(z(x))}\,g(z(x))}
    {\int dx\,dp\,e^{-(1/2)\,\gamma^{ab}\,p_a p_b - \ReS(z(x))}}. 
\end{align}
Note that the volume element can be expressed as $dx\,dp = \omega^N/N!$ 
with the symplectic 2-form $\omega\equiv dp_a\wedge dx^a$.

In Monte Carlo calculations, 
it is more convenient to rewrite everything 
in terms of the target space coordinates $z=(z^i)$. 
To do this, we introduce the momentum $\pi=(\pi^i)$ 
which is tangent to $\Sigma$: 
\begin{align}
  \pi \in T_z\Sigma \mbox{~~~with~~} \pi^i \equiv p^a E_a^i 
  \quad (p^a\equiv \gamma^{ab}\,p_b).
\end{align}
One then can show that 
the 1-form 
\begin{align}
 a\equiv \langle\pi, dz\rangle
 = {\rm Re}\,\overline{\pi^i}\,dz^i
\end{align}
can be expressed as $a=p_a dx^a$, 
and thus we find that $a$ is a symplectic potential of $\omega$, 
$\omega = da$, 
and obtain the identity 
\begin{align}
  \omega = {\rm Re}\,d\overline{\pi^i}\wedge dz^i. 
\end{align}
Furthermore, noting the identity 
\begin{align}
  \langle \pi,\pi \rangle = \gamma^{ab} p_a p_b\quad
  (\pi\in T_z\Sigma),
\end{align}
we have the following target-space representation:
\begin{align}
  \langle g(z) \rangle 
  = \frac{\int_{T\Sigma}\, \omega^N\,e^{-H(z,\pi)}\,g(z)}
    {\int_{T\Sigma}\, \omega^N\,e^{-H(z,\pi)}}.
\end{align}
Here, $T\Sigma \equiv \{(z,\pi)\,|\,z\in\Sigma,\,\pi\in T_z\Sigma\}$ 
is the tangent bundle of $\Sigma$, 
and $H(z,\pi)$ is the Hamiltonian of the form%
\footnote{ 
  A more precise expression is 
  $H(z,\bar z,\pi,\bar \pi)=(1/2)\langle \pi,\pi\rangle
  + V(z,\bar z)$, 
  but we abbreviate it as in the text 
  to simplify expressions. 
} 
\begin{align}
  H(z,\pi) = \frac{1}{2}\,\langle\pi,\pi\rangle + V(z)
\end{align}
with the (real-valued) potential 
\begin{align}
  V(z) = \ReS(z) = \frac{1}{2}\,[S(z) + \overline{S(z)}]. 
\end{align}

\subsection{Constrained molecular dynamics on $\Sigma$}
\label{sec:gt_md}

We assume that the $N$-dimensional real submanifold $\Sigma$ 
in $\bbC^N=\bbR^{2N}$ 
is specified by $N$ independent equations $\phi^r(z)=0$ 
$(r=1,\ldots,N)$
with real-valued functions $\phi^r(z)$. 
In order to define a consistent molecular dynamics on $\Sigma$, 
we consider the Hamilton dynamics 
for an action of the first-order form: 
\begin{align}
  I[z,\pi,\lambda] = 
  \int ds\,\Bigl[ \langle \pi, \circdot{z} \rangle - H(z,\pi)
  - \lambda_r\, \phi^r(z) \Bigr].
\end{align}
Here, $\circdot{z}\equiv dz/ds$, 
and $\lambda_r\in\bbR$ are Lagrange multipliers. 
Hamilton's equations are then given by%
\footnote{ 
  Note that $\overline{\partial V(z)} = (1/2)\,\overline{\partial S(z)}$ 
  because $V(z) = {\rm Re}\,S(z) = (1/2)\,[S(z)+\overline{S(z)}]$. 
} 
\begin{align}
  \circdot{z} &= \pi,
\label{gt_md1}
\\
  \circdot{\pi} &= - 2\overline{\partial V(z)}
                     - 2\lambda_r\, \overline{\partial\phi^r(z)}
\label{gt_md2}
\end{align}
with constraints 
\begin{align}
  \phi^r(z) &= 0,
\label{gt_md3}
\\
  \langle \pi, \overline{\partial\phi^r} \rangle &= 0.
\label{gt_md4}
\end{align}
One can easily show that 
the symplectic potential $a$ changes under molecular dynamics 
as $\circdot{a} = d[(1/2)\langle \pi,\pi \rangle - V(z)]$, 
from which follows $\circdot{\omega} = d\circdot{a} = 0$. 
Furthermore, noting that 
$\lambda\equiv \lambda_r\,\overline{\partial \phi^r(z)}\in N_z\Sigma$,%
\footnote{
  In fact, for any vector $v\in T_z\Sigma$, 
  we have 
  \begin{align}
    \langle \lambda,v \rangle
    = \lambda_r\, {\rm Re}\,\langle \overline{\partial\phi^r},v \rangle
    = (\lambda_r/2)\,(v\cdot\partial+\bar{v}\cdot\bar\partial)\phi^r
    = (\lambda_r/2)\,\lim_{\epsilon\to 0}(1/\epsilon)[\phi^r(z+\epsilon v)-\phi^r(z)]
    = 0.
    \nonumber
  \end{align}
} 
one can also show that $\circdot{H} = 0$.

A discretized form of Eqs.~\eqref{gt_md1}--\eqref{gt_md4} 
with step size $\Delta s$ is given by RATTLE 
\cite{Andersen:1983,Leimkuhler:1994}
of the following form 
(we rescale $\lambda\to\lambda/\Delta s$ for later convenience) 
\cite{Fujii:2013sra,Alexandru:2019,Fukuma:2019uot}:
\begin{align}
  \pi_{1/2} &= \pi - \Delta s\,\overline{\partial V(z)} - \lambda/\Delta s,
\label{gt_rattle1}
\\
  z' &= z + \Delta s\,\pi_{1/2},
\label{gt_rattle2}
\\
  \pi' &= \pi_{1/2} - \Delta s\,\overline{\partial V(z')} - \lambda'/\Delta s.
\label{gt_rattle3}
\end{align}
Here, the Lagrange multipliers $\lambda\in N_z\Sigma$ 
and $\lambda'\in N_{z'}\Sigma$ 
are determined 
such that $z'\in \Sigma$ and $\pi'\in T_{z'}\Sigma$, respectively.
One easily sees that 
the transformation $(z,\pi)\to(z',\pi')$ satisfies the reversibility.%
\footnote{ 
  If $(z,\pi)\to(z',\pi')$ is a motion, 
  so is $(z',-\pi')\to(z,-\pi)$ 
  with $\lambda$ and $\lambda'$ interchanged.
\label{fn:reversibility}
} 
Noting that $\langle \lambda,dz \rangle = 0$ 
and $\langle \lambda', dz' \rangle = 0$, 
one can also show that 
the symplectic potential $a= \langle \pi, dz \rangle$ 
transforms as follows:
\begin{align}
  a_{1/2} &\equiv \langle \pi_{1/2}, dz\rangle 
  = a - (\Delta s/2)\,dV(z),
\\
  a'_{1/2} &\equiv \langle \pi_{1/2}, dz'\rangle
  = a_{1/2} + (\Delta s/2)\,d\langle \pi_{1/2},\pi_{1/2}\rangle,
\\
  a' &\equiv \langle \pi', dz'\rangle
  = a'_{1/2} - (\Delta s/2)\,dV(z')
\nonumber
\\
  &= a + (\Delta s/2)\,d\bigl[\langle \pi_{1/2},\pi_{1/2}\rangle - V(z) - V(z')\bigr],
\end{align}
from which we find that 
this transformation is symplectic ($\omega' = \omega$) 
and thus volume-preserving  ($\omega'^N = \omega^N$). 
One can further show that this transformation preserves the Hamiltonian 
to $O(\Delta s^2)$:%
\footnote{ 
  Note that 
  $\langle \lambda,\pi \rangle = \langle \lambda',\pi' \rangle = 0$, 
  $\lambda=O(\Delta s^2)$, $\lambda'=O(\Delta s^2)$
  and $\lambda-\lambda'=O(\Delta s^3)$.
} 
\begin{align}
  H(z',\pi') = H(z,\pi) + O(\Delta s^3).
\end{align}

\subsection{Projector in GT-HMC}
\label{sec:gt_projector}

As we see in the next subsection, 
in determining $\lambda$ and $\lambda'$, 
we repeatedly project a vector $w\in T_z\bbC^N$ 
onto the tangent space $T_z\Sigma$ and the normal space $N_z\Sigma$: 
\begin{align}
  w = v + n \quad
  \bigl(v\in T_z\Sigma,~ n\in N_z\Sigma\bigr). 
\label{gt_projector}
\end{align}
This projection can be carried out 
by an iterative use of flow 
as follows \cite{Fujii:2013sra}. 
For $z\in\Sigma$ and its starting configuration $x\in\Sigma_0=\bbR^N$, 
we introduce an $\bbR$-linear map $A:\,T_{x}\bbC^N \to T_z\bbC^N$ 
which consists of three steps
(see Fig.~\ref{fig:multEF}):
\begin{figure}[tb]
  \centering
  \includegraphics[width=85mm]{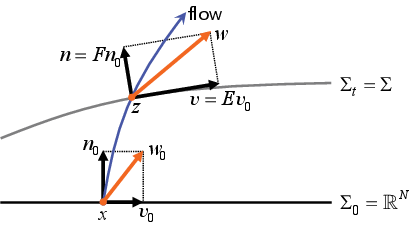}
  \caption{Linear transformation: $w_0=v_0+n_0 \mapsto w = E v_0 + F n_0$.}
  \label{fig:multEF}
\end{figure}%

\indent
\underline{(1)}~
For a given vector $w_0\in T_{x}\bbC^N$, 
decompose it into 
\begin{align}
  v_0 \equiv \frac{1}{2}\,(w_0 + \bar{w}_0) \in T_{x}\Sigma_0,\quad
  n_0 \equiv \frac{1}{2}\,(w_0 - \bar{w}_0) \in N_{x}\Sigma_0.
\end{align}
\indent
\underline{(2)}~
Integrate the following flow equations that maps 
$v_0\in T_{x}\Sigma_0$ to $v\in T_z\Sigma$ 
and $n_0\in N_{x}\Sigma_0$ to $n\in N_z\Sigma$, 
\begin{align}
  \dot{z} &= \overline{\partial S(z)} \mbox{~~with~~} z|_{t=0} = x,
\label{flow_c}
\\
  \dot{v} &= \overline{H(z)\,v} \mbox{~~with~~} v|_{t=0} = v_0,
\label{flow_ct}
\\
  \dot{n} &= - \overline{H(z)\,n}  \mbox{~~with~~} n|_{t=0} = n_0,
\label{flow_ctn}
\end{align}
where $H(z)\equiv (\partial_i \partial_j S(z))$ is the Hessian matrix.%
\footnote{ 
  Equation \eqref{flow_ct} 
  is obtained from another flow equation of type \eqref{flow_c}, 
  $(z+\epsilon v)^\centerdot = \overline{\partial S(z+\epsilon v)}$ 
  with an infinitesimal parameter $\epsilon$. 
  Then, Eq.~\eqref{flow_ctn} is deduced 
  from the condition that $\langle v,n\rangle^\centerdot = 0$.
} 
Note that $v$ and $n$ are linear in $v_0$ and $n_0$, respectively, 
and we write them as
\begin{align}
  v^i \equiv E^i_a v_0^a = (E v_0)^i,\quad
  n^i \equiv F^i_a n_0^a = (F n_0)^i.
\end{align}
\indent
\underline{(3)}~
Define an $\bbR$-linear map 
$A:\,T_{x}\bbC^N\ni w_0 \mapsto w\in T_z\bbC^N$ 
by
\begin{align}
  w = A w_0 \equiv E v_0 + F n_0 \mbox{~~for~~} w_0=v_0+n_0.
\label{gt_projection}
\end{align}

Once the map $A$ is defined, 
the decomposition \eqref{gt_projector} of a given $w\in T_z\bbC^N$ 
can be carried out as in Algorithm~\ref{alg:gt_decomposition}.
Note that, if we use an iterative method (such as BiCGStab) 
for solving $A w_0 = w$ in Step 1, 
we no longer need to carry out Step 2 and Step 3 \cite{Alexandru:2017lqr}. 
This is because in Step 1 
we repeatedly compute $E \tilde{v}_0$ and $F \tilde{n}_0$ 
for a candidate solution $\tilde{w}_0 = \tilde{v}_0 + \tilde{n}_0$, 
so that $v = E v_0$ and $n = F n_0$ are already obtained 
when the iteration is converged.
%
\begin{algorithm}[tb]
	\caption{Orthogonal decomposition of $w\in T_z\bbC^N$ into $v\in T_z\Sigma$ and $n\in N_z\Sigma$}
	\label{alg:gt_decomposition}
	
	\begin{algorithmic}[1]
		\State%
		Solve the linear problem $A w_0 = w$ with respect to $w_0$.

		\State%
		Decompose $w_0$ 
		into $v_0=(1/2)(w_0+\bar{w}_0)$ and $n_0=(1/2)(w_0-\bar{w}_0)$.

		\State%
	  Compute $v = E v_0$ and $n = F n_0$ 
	  by integrating Eqs.~\eqref{flow_c}--\eqref{flow_ctn}.		%
	\end{algorithmic}
\end{algorithm}

\subsection{RATTLE in GT-HMC}
\label{sec:gt_rattle}

The Lagrange multiplies $\lambda\in N_z\Sigma$ 
and $\lambda'\in N_{z'}\Sigma$ in Eqs.~\eqref{gt_rattle1}--\eqref{gt_rattle3} 
are determined as follows.

\subsubsection{Determination of $\lambda$}
\label{sec:gt_lambda}

\begin{figure}[tb]
  \centering
  \includegraphics[width=75mm]{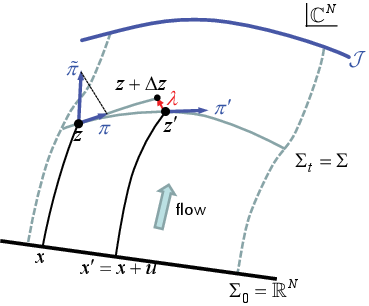}
  \caption{
  	RATTLE in GT-HMC: $(z,\pi)\to (z',\pi')$.
  	The initial momentum $\pi$ in the molecular dynamics 
  	is constructed by projecting $\tilde\pi\in T_z\bbC^N$ onto $T_z\Sigma$,
  	where $\tilde\pi$ is randomly generated with the Gaussian distribution 
  	$\propto e^{-(1/2)\tilde\pi^\dagger \tilde\pi}$.
  }
  \label{fig:gt_rattle}
\end{figure}

The condition that $z'\in \Sigma$ 
is equivalent to that $z'$ can be written as $z'=z_t(x')$ 
with $x'\in\Sigma_0$ 
(see Fig.~\ref{fig:gt_rattle}). 
Thus, finding $\lambda$ satisfying Eqs.~\eqref{gt_rattle1} and \eqref{gt_rattle2} 
for a given $z=z_t(x)\in\Sigma$ and $\pi\in T_z\Sigma$
is equivalent to finding a doublet $(u,\lambda)$ 
$(u\in T_{x}\Sigma_0,\,\lambda\in N_z\Sigma)$ 
that satisfies 
\begin{align}
  z_t(x+u) = z_t(x) + \Delta z - \lambda 
\label{gt_newton1}
\end{align}
with
\begin{align}
  \Delta z(z,\pi)\equiv \Delta s\,\pi - (\Delta s)^2\, \overline{\partial V(z)}.
\end{align}
Equation \eqref{gt_newton1} can be solved iteratively with Newton's method. 
There, one solves the following linearized equation 
in updating an approximate solution $(u,\lambda)$ 
as $u\gets u+\Delta u$  and $\lambda\gets \lambda+\Delta \lambda$:
\begin{align}
  E_{\rm new} \Delta u + \Delta \lambda = B,
\label{gt_newton2}
\end{align}
where $E_{\rm new}\equiv \partial z_t(x+u)/\partial u = (\partial z_t/\partial x)(x+u)$, 
and 
\begin{align}
  B\equiv z + \Delta z - \lambda - z_{\rm new}\in \bbC^N
\label{gt_B}
\end{align}
with $z_{\rm new}\equiv z_t(x+u)$.

Equation \eqref{gt_newton2} can be solved with direct or iterative methods  
by regarding it as a linear equation of the form $\tilde{A} X=B$
with respect to $X = (\Delta u, F^{-1}\Delta \lambda)$ 
as carried out in Ref.~\cite{Fukuma:2019uot}.%
\footnote{
  We discuss in Sec.~\ref{sec:full_newton} 
  the convergence of iteration to solve Eq.~\eqref{gt_newton1}
  when using the original equation \eqref{gt_newton2} with iterative solvers.
} 
Instead of solving Eq.~\eqref{gt_newton2}, 
we here propose to use the simplified Newton equation 
(corresponding to the fixed-point method of Ref.~\cite{Fujii:2013sra} 
for the case of the projection onto a single thimble), 
where $E_{\rm new}$ on the left hand side 
is replaced by the value at $z=z_t(x)$: 
\begin{align}
  E \Delta u + \Delta \lambda = B.
\label{gt_newton3}
\end{align}
This equation can be readily solved 
by using the projection introduced in Sect.~\ref{sec:gt_projector}. 
To see this, 
from the orthogonal decomposition 
\begin{align}
  B =& B_v + B_n 
    = E B_{0,v} + F B_{0,n}
\end{align}
with 
$B_v\in T_z\Sigma$, $B_n\in N_z\Sigma$,
$B_{0,v}\in T_{x}\Sigma_0$ and $B_{0,n}\in N_{x}\Sigma_0$,
we write $B$ to the form 
\begin{align}
  B = E B_{0,v} + B_n. 
\end{align}
Then, comparing with the left hand side of Eq.~\eqref{gt_newton3}, 
we obtain%
\footnote{
  With knowledge of the explicit form of the Jacobian $E=(E^i_a)$ 
  [requiring the computational cost of $O(N^3)$], 
  they can be written as 
  $\Delta u = {\rm Re}(E^{-1}B)$ and 
  $\Delta \lambda = F\,(i\, {\rm Im}(E^{-1}B)) = i\,E\, {\rm Im}(E^{-1}B)$.
  ${\rm Re}(E^{-1}B)$ and ${\rm Im}(E^{-1}B)$ correspond 
  to Eqs.~(3.21) and (3.22) in Ref.~\cite{Fujii:2013sra}, respectively.
} 
\begin{align}
  \Delta u = B_{0,v},\quad
  \Delta \lambda = B_n.
\end{align}
Note that, if we set the initial guess to $u=0$ and $\lambda=0$, 
then the first run in the iteration gives the following result
with respect to the decomposition 
$\Delta z = E(\Delta z)_{0,v} + (\Delta z)_n$: 
\begin{align}
	u = (\Delta z)_{0,v},\quad
	\lambda = (\Delta z)_n \quad
  \mbox{(approximate solution)}.
\label{gt_init2}
\end{align}

\subsubsection{Determination of $\lambda'$}
\label{sec:gt_lambda'}

Note that determining $\lambda'$ in Eq.~\eqref{gt_rattle3} 
such that $\pi'\in T_{z'}\Sigma$ 
is equivalent to projecting 
$\tilde\pi' \equiv \pi_{1/2} - \Delta s\,\overline{\partial V(z')}$ 
onto $T_{z'}\Sigma$. 
Thus, $\pi'$ is simply obtained from the decomposition 
$\tilde\pi' = \tilde\pi'_v + \tilde\pi'_n$ 
at $z'$ as $\pi' = \tilde\pi'_v$.

Molecular dynamics from a configuration $(z,\pi) \in T\Sigma$  
is summarized in Algorithm \ref{alg:gt_rattle}. 
%
\begin{algorithm}[tb]
\caption{Simplified RATTLE $(z,\pi)\to(z',\pi')$ in GT-HMC}
\label{alg:gt_rattle}

\begin{algorithmic}[1]
\State%
  Compute $\Delta z = \Delta s\,\pi - (\Delta s)^2\,\overline{\partial V(z)}$

\State%
  Set $u = 0$ and $\lambda = 0$
\For{$k=0,1,\ldots$}%
  \State%
    Compute $z_{\rm new}=z_t(x+u)$ and set $B = z + \Delta z - \lambda - z_{\rm new}$
  \If{$|B|$ is small} 
    \State \textbf{break}
  \EndIf
  \State%
    Decompose $B$ into $B = E B_{0,v} + B_n$
  \State%
    Set $\Delta u = B_{0,v}$ and $\Delta\lambda = B_n$
  \State%
    $u \leftarrow u+ \Delta u$ and $\lambda \leftarrow \lambda + \Delta \lambda$
\EndFor
\State%
  Set $z' = z_{\rm new}$ and 
  $\tilde{\pi}'
  = \pi - \Delta s\,[\overline{\partial V(z)} + \overline{\partial V(z')}]
    - \lambda/\Delta s$
\State%
  Decompose $\tilde\pi'\in T_{z'}\bbC^N$ 
  into $\tilde\pi' = \tilde\pi'_v + \tilde\pi'_n$ 
  and set $\pi' = \tilde\pi'_v$
\end{algorithmic}
\end{algorithm}

\subsection{Summary of GT-HMC}
\label{sec:gt_summary}

We summarize in Algorithm \ref{alg:gt_hmc} 
the GT-HMC algorithm for updating a configuration $z\in\Sigma$ 
with the RATTLE of Algorithm~\ref{alg:gt_rattle}.%
\footnote{ 
  One needs to compute the phase of the Jacobian determinant, 
  $dz/|dz| = \det E/|\det E|$, upon measurement, 
  of which the direct computation costs $O(N^3)$. 
  However, the phase can be evaluated 
  by using a stochastic estimator, 
  for which the computational cost is reduced to $O(N\times N_R)$, 
  where $N_R$ is the number of independent Gaussian noise fields 
  \cite{Cristoforetti:2014gsa}.
} 

%
\begin{algorithm}[tb]
\caption{GT-HMC}
\label{alg:gt_hmc}

\begin{algorithmic}[1]
\State%
  Given $z\in\Sigma$, 
  generate $\tilde\pi\in T_z\bbC^N$ from the distribution 
  $\propto e^{-\tilde\pi^\dagger \tilde\pi/2}$
\State%
  Decompose $\tilde\pi$ into $\tilde\pi = \tilde\pi_v + \tilde\pi_n$ 
  and set $\pi = \tilde\pi_v$
\State%
  Repeat the RATTLE of Algorithm \ref{alg:gt_rattle} 
  to obtain $(z,\pi) \to (z',\pi')$ 
\State%
  Compute $\Delta H\equiv H(z',\pi') - H(z,\pi)$ 
  and accept $(z',\pi')$ as a new configuration 
  with probability $\min(1,e^{-\Delta H})$, 
  otherwise use $(z,\pi)$ again as a new configuration
\end{algorithmic}
\end{algorithm}

\section{Worldvolume Hybrid Monte Carlo (WV-HMC)}
\label{sec:wv}

We explain the basics of WV-HMC \cite{Fukuma:2020fez} 
and propose its simplified algorithm. 
For convenience of the reader who reads only this section, 
the presentation is made 
in a completely parallel way to that for GT-HMC 
without worrying about repetition.

\subsection{Path-integral form for WV-HMC}
\label{sec:wv_pi}

We restart from the expression \eqref{vev_Sigma}:
\begin{align}
  \langle \calO \rangle = 
  \frac{\int_{\Sigma_t} dz_t\,e^{-S(z_t)}\,\calO(z_t)}
  {\int_{\Sigma_t} dz_t\,e^{-S(z_t)}},
\end{align}
where we have denoted the configurations by $z_t$ 
(instead of $z$)
to stress that they live on $\Sigma_t$. 
Since the numerator and the denominator are both independent of $t$ 
(Cauchy's theorem), 
they can be averaged over flow time $t$ 
with an arbitrary weight $e^{-W(t)}$, 
leading to the expression \cite{Fukuma:2020fez}
\begin{align}
  \langle \calO \rangle 
  &= 
  \frac{\int dt\,e^{-W(t)}\,\int_{\Sigma_t} dz_t\,e^{-S(z_t)}\,\calO(z_t)}
  {\int dt\,e^{-W(t)}\,\int_{\Sigma_t} dz_t\,e^{-S(z_t)}}
\nonumber\\
  &\equiv
  \frac{\int_{\calR} dt\,dz_t\,e^{-S(z_t)-W(t)}\,\calO(z_t)}
  {\int_{\calR} dt\,dz_t\,e^{-S(z_t)-W(t)}}
  \equiv \frac{Z_{\calO}}{Z}. 
\label{wv_pi1}
\end{align}
The $(N+1)$-dimensional integration region $\calR$ is defined 
by $\calR \equiv \{z_t\in\Sigma_t \,|\,t\in \bbR\}$, 
which we refer to the \emph{worldvolume}, 
by regarding it as an orbit of an integration surface $\Sigma_t$
in the target space $\bbC^N=\bbR^{2N}$. 
The extension of $\calR$ in the flow time direction 
can be effectively constrained to a finite interval $[T_0,T_1]$ 
by adjusting the functional form of $W(t)$. 
The function $W(t)$ has another role 
to lift configurations upwards (positive flow direction) 
so that they are distributed almost equally over different flow times. 
In fact, the force of molecular dynamics  
exerts configurations in the direction opposite to the flow 
[see, e.g., 
Eq.~\eqref{gt_md2} 
with $-2\overline{\partial V(z)} = -\,\overline{\partial S(z)}$
for the case of GT-HMC],
and thus configurations have a tendency to precipitate towards the bottom (near $\Sigma_0$) 
if nothing is done. 
A possible form of $W(t)$ is given in Sect.~\ref{sec:wv_boundary} 
(see Ref.~\cite{fn1} for a more detailed study).

Since multimodality becomes more severe 
at larger flow times, 
we take the lower bound $T_0$ to be small enough 
such that there is no ergodicity problem on $\Sigma_{T_0}$.%
\footnote{ 
  When the system already has an ergodicity problem 
  on the original integration surface $\Sigma_{t=0}$, 
  we further implement other algorithms to reduce the problem 
  or use $T_0$ of a negative value \cite{Fukuma:2017fjq}.
} 
The upper bound $T_1$ is chosen 
such that oscillatory integrals are sufficiently tamed there.%
\footnote{ 
  By using the GT-HMC, 
  one can set a criterion for the choice of $T_1$, e.g., 
  that the average phase factor 
  $\langle e^{-i\ImS(z)}\,dz/|dz| \rangle_{\Sigma_{T_1}}$
  is not zero at least to two standard deviations.  
} 
After global equilibrium is well established over $\calR$, 
we estimate the expectation value $\langle \mathcal{O} \rangle$ 
with sample averages 
using the configurations taken from a subinterval $[\tilde{T}_0,\tilde{T}_1]$ 
($T_0 \leq \tilde{T}_0 < \tilde{T}_1 \leq T_1$), 
which is free from both of the sign problem at $t\sim T_0$
and the possible complicated geometry at $t\sim T_1$.%
\footnote{ 
  The subinterval for estimation, $[\tilde{T}_0,\tilde{T}_1]$, 
  is determined by the condition
  that the estimate of $\calO$ only varies within small statistical errors 
  against small changes of the subinterval \cite{Fukuma:2020fez}. 
  The set of configurations in the subregion 
  $\tilde\calR \equiv \{z\in\Sigma_t\,|\,t\in [\tilde T_0,\tilde T_1]\}$ 
  can also be regarded as a Markov chain, 
  so that the standard statistical analysis method (such as Jackknife) 
  can also be applied \cite{Fukuma:2021aoo}.
} 
 
With local coordinates 
$x=(x^a)$ for $\Sigma_t$ (see footnote~\ref{fn:coords}), 
we introduce those of $\calR$ as 
$\hat{x}=(\hat{x}^\mu)=(\hat{x}^0=t,\hat{x}^a=x^a)$. 
Then, the induced metric on $\calR$, 
$d\hat{s}^2 = \hat\gamma_{\mu\nu}\,d\hat{x}^\mu d\hat{x}^\nu
\equiv |dz(\hat{x})|^2$, 
takes the following form \cite{Fukuma:2020fez}:
\begin{align}
  d\hat{s}^2 
  &= |(\partial_t z^i)\,dt + (\partial_{x^a}z^i)\,dx^a|^2
  = |\xi^i dt + E_a^i dx^a|^2
  = |\xi^i_n dt + (E_a^i dx^a + \xi_v^i dt)|^2
\nonumber
\\
  &= \alpha^2\,dt^2 + \gamma_{ab}\,(dx^a+\beta^a dt)\,(dx^b+\beta^b dt)
\end{align}
with the induced metric $\gamma_{ab}$ on $\Sigma_t$, the shift $\beta^a$ 
and the lapse $\alpha\,(>0)$ 
given by
\begin{align}
  \gamma_{ab} &= \langle E_a, E_b \rangle,
\\
  \beta^a &= \gamma^{ab}\,\langle \xi, E_b \rangle,
\\
  \alpha^2 &= \langle \xi_n,\xi_n\rangle.
\end{align}
Here, 
$E_a=(E^i_a=\partial z^i/\partial x^a)$ form a basis of $T_z\Sigma_t$, 
and $\xi=\overline{\partial S}$ is the flow vector, 
which is decomposed into the tangential and the normal components 
as $\xi=\xi_v + \xi_n$ [see Eq.~\eqref{gt_projector}]. 
Note that 
$\xi_v=E_a\,\beta^a \in T_z\Sigma_t\,(\subset T_z\calR)$
and $\xi_n\in N_z\Sigma_t\cap T_z\calR$. 
The invariant volume element on $\calR$ is then given by 
\begin{align}
  |dz|_\calR \equiv \sqrt{\hat\gamma}\,d\hat{x} = \alpha\sqrt{\gamma}\,dt\,dx
  = \alpha\,|dz_t|\,dt,
\end{align}
and $Z_\calO$ in Eq.~\eqref{wv_pi1} can be written as 
\begin{align}
  Z_\calO = \int_\calR |dz|_\calR\,e^{-V(z)}\,\calF(z)\,\calO(z)
\end{align}
with
\begin{align}
  V(z) &\equiv \ReS(z) + W(t(z)),
\\
  \calF(z) &\equiv \frac{dt\,dz_t}{|dz|_\calR}\,e^{-i\,\ImS(z)}
  = \alpha^{-1}\,\frac{dz_t}{|dz_t|}\,e^{-i\,\ImS(z)}
  = \alpha^{-1}\,\frac{\det E}{|\det E\,|}\,e^{-i\,\ImS(z)}.
\end{align}
Thus, by defining the reweighted average $\langle\cdots\rangle_\calR$ on $\calR$ by 
\begin{align}
  \langle g(z) \rangle_\calR \equiv 
  \frac{\int_\calR |dz|_\calR\,e^{-V(z)}\,g(z)}
  {\int_\calR |dz|_\calR\,e^{-V(z)}},
\end{align}
the expectation value \eqref{wv_pi1} is expressed 
as a ratio of the reweighted averages: 
\begin{align}
  \langle \calO \rangle 
  = \frac{\langle \calF(z)\,\calO(z) \rangle_\calR}{\langle \calF(z) \rangle_\calR}.
\end{align}

Similarly to the GT-HMC algorithm, 
the reweighted averages $\langle \cdots \rangle_\calR$  
can be written as a path integral over the phase space 
by rewriting the measure 
$|dz|_\calR=\sqrt{\hat\gamma}\,d\hat{x}$ 
to the form 
\begin{align}
  |dz|_\calR = \sqrt{\hat\gamma}\,d\hat{x} 
  \propto d\hat{x}\,d\hat{p}\,e^{-(1/2)\,\hat\gamma^{\mu\nu}\,\hat{p}_\mu \hat{p}_\nu}, 
\end{align}
where $d\hat{x}\,d\hat{p}\equiv \prod_\mu \bigl( d\hat{x}^\mu d\hat{p}_\mu \bigr)$ 
is the volume element of the phase space of $\calR$ 
and $(\hat\gamma^{\mu\nu})\equiv (\hat\gamma_{\mu\nu})^{-1}$. 
We thus obtain the phase-space path integral  
in the parameter-space representation: 
\begin{align}
  \langle g(z) \rangle_\calR 
  = \frac{\int d\hat{x}\,d\hat{p}\,
    e^{-(1/2)\,\hat\gamma^{\mu\nu}\,\hat{p}_\mu \hat{p}_\nu
      - V(z(\hat{x}))}\,g(z(\hat{x}))}
    {\int d\hat{x}\,d\hat{p}\,e^{-(1/2)\,\hat\gamma^{\mu\nu}\,\hat{p}_\mu \hat{p}_\nu
      - V(z(\hat{x}))}}. 
\end{align}
Note that the volume element can be expressed as 
$d\hat{x}\,d\hat{p} = \hat\omega^{N+1}/(N+1)!$ 
with the symplectic 2-form 
$\hat\omega \equiv d\hat{p}_\mu\wedge d\hat{x}^\mu$. 

In Monte Carlo calculations, 
it is more convenient to rewrite everything 
in terms of the target space coordinates $z=(z^i)$. 
To do this, we introduce the momentum $\pi=(\pi^i)$ 
which is tangent to $\calR$: 
\begin{align}
  \pi \in T_z\calR \mbox{~~~with~~} \pi^i \equiv \hat{p}^\mu \hat{E}_\mu^i 
  \quad 
  (\hat{p}^\mu\equiv \hat\gamma^{\mu\nu}\,\hat{p}_\nu,~
   \hat{E}_\mu^i\equiv \partial z^i/\partial \hat{x}^\mu).
\end{align}
One then can show that 
the 1-form 
\begin{align}
 \hat{a}\equiv \langle\pi, dz\rangle
 = {\rm Re}\,\overline{\pi^i}\,dz^i
\end{align}
can be expressed as $\hat{a}=\hat{p}_\mu d\hat{x}^\mu$, 
and thus we find that $\hat{a}$ is a symplectic potential of $\hat\omega$, 
$\hat\omega = d\hat a$, 
and obtain the identity 
\begin{align}
  \hat\omega = {\rm Re}\,d\overline{\pi^i}\wedge dz^i. 
\end{align}
Furthermore, noting the identity 
\begin{align}
  \langle \pi,\pi \rangle = \hat\gamma^{\mu\nu} \hat{p}_\mu \hat{p}_\nu 
  \quad (\pi\in T_z\calR),
\end{align}
we have the following target-space representation:
\begin{align}
  \langle g(z) \rangle_\calR 
  = \frac{\int_{T\calR}\, \hat\omega^{N+1}\,e^{-H(z,\pi)}\,g(z)}
    {\int_{T\calR}\, \hat\omega^{N+1}\,e^{-H(z,\pi)}}.
\end{align}
Here, $T\calR \equiv \{(z,\pi)\,|\,z\in\calR,\,\pi\in T_z\calR\}$ 
is the tangent bundle of $\calR$, 
and $H(z,\pi)$ is the Hamiltonian of the form 
\begin{align}
  H(z,\pi) = \frac{1}{2}\,\langle\pi,\pi\rangle + V(z)
\end{align}
with the (real-valued) potential 
\begin{align}
  V(z) = \ReS(z) + W(t(z)). 
\end{align}

\subsection{Constrained molecular dynamics on $\calR$}
\label{sec:wv_md}

In parallel to discussions for GT-HMC, 
the RATTLE \cite{Andersen:1983,Leimkuhler:1994} for WV-HMC is given as follows 
\cite{Fukuma:2020fez}: 
\begin{align}
  \pi_{1/2} &= \pi - \Delta s\,\overline{\partial V(z)} - \lambda/\Delta s,
\label{wv_rattle1}
\\
  z' &= z + \Delta s\,\pi_{1/2},
\label{wv_rattle2}
\\
  \pi' &= \pi_{1/2} - \Delta s\,\overline{\partial V(z')} - \lambda'/\Delta s.
\label{wv_rattle3}
\end{align}
Here, the Lagrange multipliers $\lambda\in N_z\calR$ 
and $\lambda'\in N_{z'}\calR$ 
are determined 
such that $z'\in \calR$ and $\pi'\in T_{z'}\calR$, respectively.
This transformation satisfies the reversibility 
as in footnote~\ref{fn:reversibility}. 
One can also show that this transformation is symplectic 
($\hat\omega' = \hat\omega$) 
and thus volume-preserving  ($\hat\omega'^{N+1} = \hat\omega^{N+1}$). 
One can further show that it preserves the Hamiltonian 
to $O(\Delta s^2)$: $H(z',\pi') = H(z,\pi) + O(\Delta s^3)$.

The gradient of the potential, $\overline{\partial V(z)}$, 
can be set to the form \cite{Fukuma:2020fez}
\begin{align}
  \overline{\partial V(z)} = 
  \frac{1}{2}\,\Bigl[ \xi + \frac{W'(t)}
  {\langle \xi_n,\xi_n \rangle}\,\xi_n \Bigr].
\label{wv_force}
\end{align}
A simplified proof is given in Appendix~\ref{sec:wv_force}.

\subsection{Projector in WV-HMC}
\label{sec:wv_projector}

As we see in the next subsection, 
in determining $\lambda$ and $\lambda'$, 
we repeatedly project a vector $w\in T_z\bbC^N$ 
onto the tangent space $T_z\calR$ and the normal space $N_z\calR$: 
\begin{align}
  w = w_\parallel + w_\perp \quad
  \bigl(w_\parallel\in T_z\calR,~ w_\perp\in N_z\calR\bigr). 
\label{wv_projector1}
\end{align}
This decomposition can be carried out 
by using the projection for $\Sigma=\Sigma_t$ 
[see Eq.~\eqref{gt_projector} and Algorithm~\ref{alg:gt_decomposition}].
In fact, 
let us decompose the vectors $w$ and $\xi$ into
\begin{align}
  w &= w_v + w_n \quad \bigl(w_v\in T_z\Sigma_t,~ w_n\in N_z\Sigma_t\bigr),
\\
  \xi &= \xi_v + \xi_n \quad \bigl(\xi_v\in T_z\Sigma_t,~ \xi_n\in N_z\Sigma_t\bigr).
\end{align}
Then, $w_\parallel$ and $w_\perp$ are given by%
\footnote{ 
  One can easily show that 
  $\langle \xi_n,w_\perp \rangle=0$
  and $\langle w_\parallel, w_\perp \rangle = 0$. 
} 
\begin{align}
  w_\parallel = w_v + c\,\xi_n,\quad
  w_\perp = w_n - c\,\xi_n
\end{align}
with 
\begin{align}
	c = \frac{\langle \xi_n, w_n\rangle}{\langle \xi_n,\xi_n \rangle}.
\end{align}

\subsection{RATTLE in WV-HMC}
\label{sec:wv_rattle}

The Lagrange multiplies $\lambda\in N_z\calR$ 
and $\lambda'\in N_{z'}\calR$ in Eqs.~\eqref{wv_rattle1}--\eqref{wv_rattle3} 
are determined as follows.

\subsubsection{Determination of $\lambda$}
\label{sec:wv_lambda}

\begin{figure}[tb]
  \centering
  \includegraphics[width=75mm]{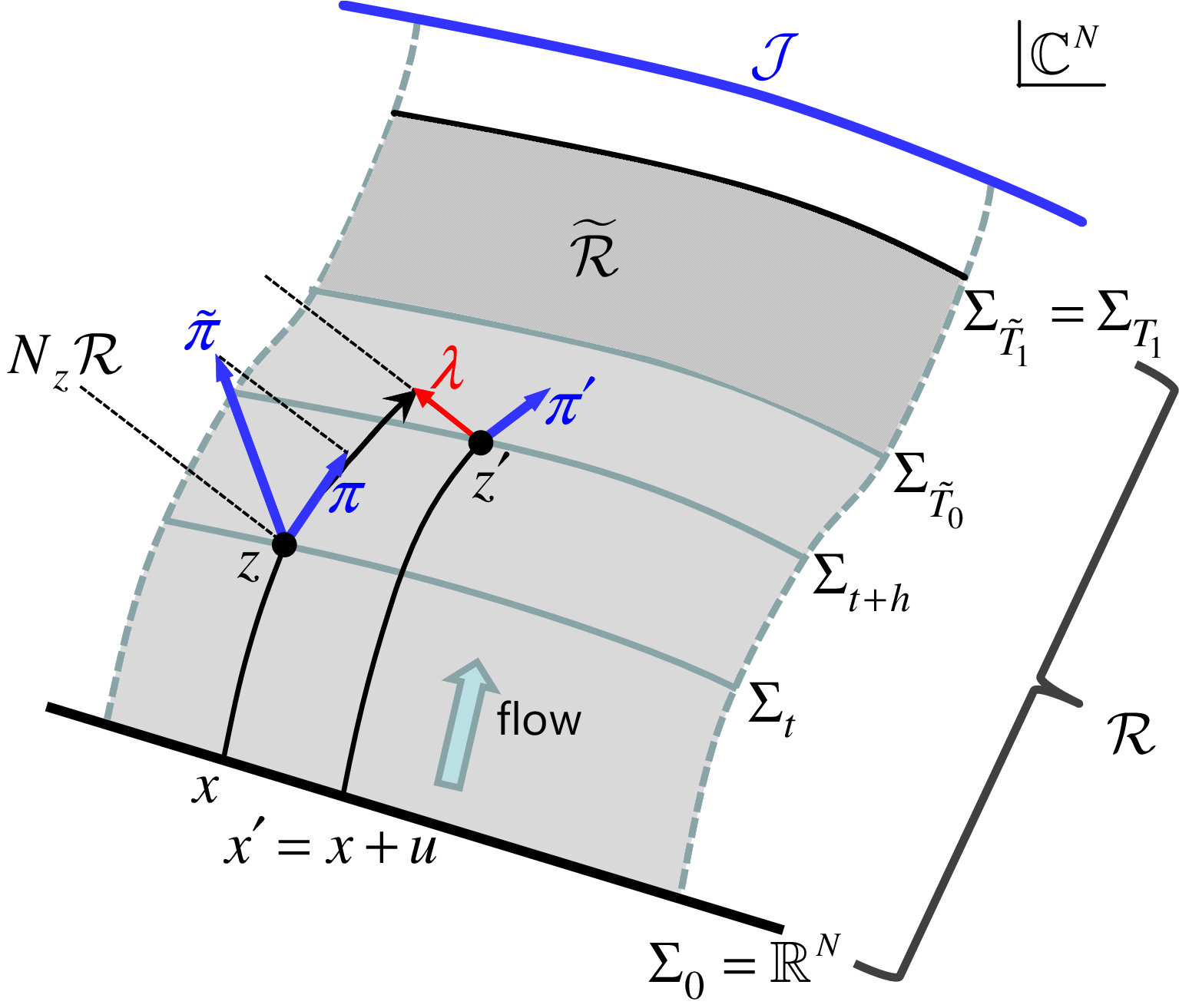}
  \caption{
  	RATTLE in WV-HMC: $(z,\pi)\to (z',\pi')$. 
  	The initial momentum $\pi$ in the molecular dynamics 
  	is constructed by projecting $\tilde\pi\in T_z\bbC^N$ onto $T_z\calR$, 
  	where $\tilde\pi$ is randomly generated 
  	with the Gaussian distribution 
  	$\propto e^{-(1/2)\tilde\pi\dagger \tilde\pi}$. 
  }
  \label{fig:wv_rattle}
\end{figure}

The condition that $z'\in \calR$ 
is equivalent to that $z'$ can be written as $z'=z_{t'}(x')$ 
with some $t'\in\bbR$ and $x'\in\Sigma_0$ 
(see Fig.~\ref{fig:wv_rattle}). 
Thus, finding $\lambda$ satisfying Eqs.~\eqref{wv_rattle1} and \eqref{wv_rattle2} 
for a given $z=z_t(x)\in\calR$ and $\pi\in T_z\calR$
is equivalent to finding a triplet $(h,u,\lambda)$ 
$(h\in\bbR,\,u\in T_{x}\Sigma_0,\,\lambda\in N_z\calR)$ 
that satisfies 
\begin{align}
  z_{t+h}(x+u) = z_t(x) + \Delta z - \lambda 
\label{wv_newton1}
\end{align}
with
\begin{align}
  \Delta z(z,\pi)\equiv \Delta s\,\pi - (\Delta s)^2\, \overline{\partial V(z)}.
\end{align}
Equation \eqref{wv_newton1} can be solved iteratively with Newton's method. 
There, one solves the following linearized equation 
in updating an approximate solution $(h,u,\lambda)$ 
as $h\gets h+\Delta h$, $u\gets u+\Delta u$ and $\lambda\gets \lambda+\Delta\lambda$:
\begin{align}
  \xi_{\rm new} \Delta h + E_{\rm new} \Delta u + \Delta \lambda = B,
\label{wv_newton2}
\end{align}
where $\xi_{\rm new}\equiv \partial z_{t+h}(x+u)/\partial h 
 = (\partial z_{t+h}/\partial t)(x+u)$, 
$E_{\rm new}\equiv \partial z_{t+h}(x+u)/\partial u
 = (\partial z_{t+h}/\partial x)(x+u)$,
and
\begin{align}
  B\equiv z + \Delta z - \lambda - z_{\rm new}\in \bbC^N
\end{align}
with $z_{\rm new}\equiv z_{t+h}(x+u)$.

Equation \eqref{wv_newton2} is proposed in Ref.~\cite{Fukuma:2020fez}, 
and can be solved with direct or iterative methods 
by regarding it as a linear equation of the form $AX=B$
with respect to $X = (\Delta h, \Delta u, \Delta \lambda)$. 
Instead of solving Eq.~\eqref{wv_newton2}, 
we here propose to use the simplified Newton equation 
(corresponding to the fixed-point method of Ref.~\cite{Fujii:2013sra} 
for the case of the projection onto a single thimble), 
where $\xi_{\rm new}$ and $E_{\rm new}$ on the left hand side 
are replaced by the values at $z=z_t(x)$: 
\begin{align}
  \xi \Delta h + E \Delta u + \Delta \lambda = B.
\label{wv_newton3}
\end{align}
This equation can be readily solved 
by using the projection introduced in Sect.~\ref{sec:wv_projector}. 
To see this, 
we introduce the decomposition of $\xi$ and $B$ as 
\begin{align}
  \xi &= \xi_\parallel
\nonumber
\\
  &= E \xi_{0,v} + \xi_n,
\label{xi_decomp}
\\
  B &= B_\parallel + B_\perp
  = (B_v + c_B\,\xi_n) + (B_n - c_B\,\xi_n)
\label{wv_rhs}
\nonumber
\\
  &= E B_{0,v} + c_B\,\xi_n + (B_n - c_B\,\xi_n)
\end{align}
with
\begin{align}
  c_B \equiv \frac{\langle B, \xi_n \rangle}{\langle \xi_n, \xi_n \rangle}.
\end{align}
Meanwhile, 
the left hand side of Eq.~\eqref{wv_newton3} is decomposed as 
\begin{align}
  \xi \Delta h + E \Delta u + \Delta \lambda
  &= (E \xi_{0,v} + \xi_n)\,\Delta h + E \Delta u + \Delta \lambda
\nonumber
\\
  &= E(\xi_{0,v}\,\Delta h + \Delta u) + \xi_n\,\Delta h + \Delta \lambda.  
\label{wv_lhs}
\end{align}
Comparing Eqs.~\eqref{wv_rhs} and \eqref{wv_lhs}, 
we find 
\begin{align}
  \xi_{0,v}\,\Delta h + \Delta u &= B_{0,v},
\\
  \Delta h &= c_B,
\\
  \Delta \lambda &= B_n - c_B\,\xi_n,
\end{align}
or equivalently,%
\footnote{
  If we set the initial guess to $h=0$, $u=0$ and $\lambda=0$, 
  then the first run in the iteration gives the result
  \begin{align}
	h = c_{\Delta z},\quad
	u = (\Delta z)_{0,v} - c_{\Delta z}\,\xi_{0,v},\quad
	\lambda = (\Delta z)_n - c_{\Delta z}\,\xi_n
  \quad
  \mbox{(approximate solution)}
  \nonumber
  \end{align}
  with respect to the decompositions $\xi = E \xi_{0,v} + \xi_n$ 
  [Eq.~\eqref{xi_decomp}] 
  and 
  \begin{align}
	\Delta z = E (\Delta z)_{0,v} + c_{\Delta z}\,\xi_n 
	+ [(\Delta z)_n - c_{\Delta z}\,\xi_n]
	\quad
	\Bigl(
	c_{\Delta z} \equiv 
	\frac{\langle \Delta z, \xi_n \rangle}{\langle \xi_n, \xi_n \rangle} 
	\Bigr).
  \nonumber
  \end{align}
} 
\begin{align}
  \Delta h = c_B,
  \quad
  \Delta u = B_{0,v} - c_B\,\xi_{0,v},
  \quad
  \Delta \lambda = B_n - c_B\,\xi_n.
\end{align}

\subsubsection{Determination of $\lambda'$}
\label{sec:wv_lambda'}

Note that determining $\lambda'$ in Eq.~\eqref{wv_rattle3} 
such that $\pi'\in T_{z'}\calR$ 
is equivalent to projecting 
$\tilde\pi' \equiv \pi_{1/2} - \Delta s\,\overline{\partial V(z')}$ 
onto $T_{z'}\calR$. 
Thus, $\pi'$ is simply obtained from the decomposition 
$\tilde\pi' = \tilde\pi'_\parallel + \tilde\pi'_\perp$ 
at $z'$ as $\pi' = \tilde\pi'_\parallel$.

Molecular dynamics from a configuration $(z,\pi) \in T\calR$  
is summarized in Algorithm \ref{alg:wv_rattle}. 
%
\begin{algorithm}[tb]
\caption{Simplified RATTLE $(z,\pi)\to(z',\pi')$ in WV-HMC}
\label{alg:wv_rattle}

\begin{algorithmic}[1]
\State%
  Compute $\xi=\overline{\partial S(z)}$ and 
  $\Delta z = \Delta s\,\pi - (\Delta s)^2\,\overline{\partial V(z)}$
\State%
  Set $h =0$,
  $u = 0$ and 
  $\lambda = 0$
\For{$k=0,1,\ldots$}%
  \State%
    Compute $z_{\rm new}=z_{t+h}(x+u)$ 
    and set $B = z + \Delta z - \lambda - z_{\rm new}$
  \If{$|B|$ is small} 
    \State \textbf{break}
  \EndIf
  \State%
    Decompose $B$ into $B = E B_{0,v} + B_n$ %
  \State%
    Set $\Delta h = c_B$, $\Delta u = B_{0,v} - c_B\,\xi_{0,v}$ %
    and $\Delta\lambda = B_n - c_B\,\xi_n$ 
    with $c_B = \langle B, \xi_n \rangle / \langle \xi_n, \xi_n \rangle$  %
  \State%
    $h \leftarrow h+ \Delta h$, $u \leftarrow u+ \Delta u$ 
    and $\lambda \leftarrow \lambda + \Delta \lambda$
\EndFor
\State%
  Set $z' = z_{\rm new}$ and 
  $\tilde{\pi}'
  = \pi - \Delta s\,[\overline{\partial V(z)} + \overline{\partial V(z')}]
    - \lambda/\Delta s$
\State%
  Decompose $\tilde\pi'\in T_{z'}\bbC^N$ 
  into $\tilde\pi' = \tilde\pi'_\parallel + \tilde\pi'_\perp$ 
  and set $\pi' = \tilde\pi'_\parallel$
\end{algorithmic}
\end{algorithm}

\subsection{Treatment of the boundary}
\label{sec:wv_boundary}

We require configurations in molecular dynamics 
to be confined in the region $T_0\lesssim t\lesssim T_1$. 
This can be realized by adjusting the function $W(t)$, 
whose possible form, e.g., is (see Ref.~\cite{fn1})%
\footnote{ 
  $\gamma\,(>0)$ is the gradient of the tilt 
  that lifts configurations upwards (positive flow direction). 
  If this simple form is not enough 
  for configurations to distribute almost equally 
  over different flow times, 
  one may resort to the multicanonical algorithm to tune $W(t)$, 
  as employed in Ref.~\cite{Fukuma:2020fez}.
} 
\begin{align}
  W(t) = 
  \begin{cases}  
    -\,\gamma(t-T_0) + c_0\,\bigl(e^{(t-T_0)^2/2d_0^2} - 1\bigr) & (t < T_0) \\
    -\,\gamma(t-T_0)                            & (T_0 \leq t \leq T_1) \\
    -\,\gamma(t-T_0) + c_1\,\bigl(e^{(t-T_1)^2/2d_1^2} - 1\bigr) & (t > T_1). \\
  \end{cases}
\end{align}
Configurations then bounce off the walls 
placed at the lower boundary ($t = T_0)$ 
and at the upper boundary ($t = T_1$) 
with penetration depths $d_0$ and $d_1$, respectively 
($c_0$ and $c_1$ correspond to the heights 
at $t=T_0-d_0$ and $t=T_1+d_1$ 
with the gradients $-\gamma - c_0/d_0$ and $-\gamma + c_1/d_1$). 
However, with a finite step size $\Delta s$, 
some configurations may penetrate the wall so deeply 
that the resulting large repulsive force 
$-W'(t)\,\overline{\partial t(z)}$ in $-\overline{\partial V(z)}$ 
can lower the numerical precision. 
The simplest solution to this issue, 
keeping (1) exact volume preservation, 
(2) exact reversibility, and (3) approximate energy conservation, 
is to let such a configuration go back the way it just comes \cite{Fukuma:2020fez}. 
The algorithm will take the form of Algorithm~\ref{alg:wv_boundary}.  
%
\begin{algorithm}[tb]
\caption{Molecular dynamics step $(z,\pi)\to(z',\pi')$ in WV-HMC with boundary}
\label{alg:wv_boundary}

\begin{algorithmic}[1]
\State%
  For a given $(z,\pi)$ with $z=z_t(x)$, 
  compute a trial molecular dynamics step 
  $(z,\pi) \to (\tilde z,\tilde\pi)$ with $\tilde z=z_{\tilde t}(\tilde x)$
  using the RATTLE of Algorithm~\ref{alg:wv_rattle}
\If{$\tilde t<T_0-d_0$ or $\tilde t>T_1+d_1$}%
  \State%
    Set $z'=z$ and $\pi'=-\pi$
\Else%
  \State%
    Set $z'=\tilde z$ and $\pi'= \tilde\pi$
\EndIf
\end{algorithmic}
\end{algorithm}

\subsection{Summary of WV-HMC}
\label{sec:wv_summary}

We summarize in Algorithm \ref{alg:wv_hmc} 
the WV-HMC method with the simplified RATTLE algorithm. 
%
\begin{algorithm}[tb]
\caption{WV-HMC}
\label{alg:wv_hmc}

\begin{algorithmic}[1]
\State%
  Given $z\in\calR$, 
  generate $\tilde\pi\in T_z\bbC^N$ from the distribution 
  $\propto e^{-\tilde\pi^\dagger \tilde\pi/2}$
\State%
  Decompose $\tilde\pi$ into 
  $\tilde\pi = \tilde\pi_\parallel + \tilde\pi_\perp$ 
  and set $\pi = \tilde\pi_\parallel$
\State%
  Repeat the RATTLE of Algorithms~\ref{alg:wv_rattle} and \ref{alg:wv_boundary}
  to obtain $(z,\pi) \to (z',\pi')$ 
\State%
  Compute $\Delta H\equiv H(z',\pi') - H(z,\pi)$ 
  and accept $(z',\pi')$ as a new configuration 
  with probability $\min(1,e^{-\Delta H})$, 
  otherwise use $(z,\pi)$ again as a new configuration
\end{algorithmic}
\end{algorithm}

\section{Numerical tests}
\label{sec:test}

In this section, 
we perform a numerical test 
for the convergence of the simplified RATTLE algorithm 
that uses the simplified Newton method (fixed-point method) 
combined with iterative solvers for orthogonal decompositions of vectors, 
and show that the convergence depends on the system size only weakly. 
We give a discussion only for the GT-HMC algorithm. 
This is because the comparison of computational costs for different system sizes 
can be made more precisely 
if the flow time is fixed.
Note that the computational cost with the WV-HMC algorithm 
are generally smaller than that with the GT-HMC algorithm. 
In fact, the computational costs for GT-HMC and WV-HMC are almost the same 
for a fixed flow time, 
and the flow times appearing in WV-HMC 
are smaller than the flow time set in GT-HMC. 
We also demonstrate 
that the computational cost of the simplified RATTLE algorithm 
is much less than the original algorithm \cite{Fukuma:2019uot} 
that uses the standard (non-simplified) Newton method.

\subsection{Convergence and numerical cost of the simplified RATTLE algorithm}
\label{sec:full_newton}

We consider the complex scalar field theory at finite density, 
whose lattice action \cite{Aarts:2008wh} is given by 
\begin{align}
  S(\phi) = \sum_n\Bigl[
  (2d+m^2)\,|\phi_n|^2 + \lambda\,|\phi_n|^4
  - \sum_{\nu=0}^{d-1}\bigl(e^{\mu\,\delta_{\nu,0}}\,\bar{\phi}_n \phi_{n+\nu}
    + e^{-\mu\,\delta_{\nu,0}}\,\bar{\phi}_{n+\nu} \phi_n \bigr) \Bigr].
\end{align}
Here, $\phi_n$ is the value at site $n$ of a complex field $\phi$ 
living on a $d$-dimensional periodic square lattice of size $L^d$, 
and $\mu$ is the chemical potential 
that makes the action complex-valued. 
We decompose $\phi_n$ into the real and imaginary parts as 
$\phi_n = \phi_{R,n} + i\,\phi_{I,n}$. 
Then, the set $\Sigma_0 \equiv \{x=(\phi_{R},\phi_{I})\}$ is the configuration space 
whose real dimension is $N=2\,L^d$, 
and we apply the WV/GT-HMC method 
following the prescriptions given in the preceding sections.%
\footnote{ 
  A detailed study of this model is given in Ref.~\cite{fn1}.
} 
In performing numerical tests, 
we set the physical parameters 
to $d=2$, $m=0.1$, $\lambda=1.0$, $\mu=0.5$, 
and vary the lattice size $L^2$ from $16^2$ to $512^2$. 
The flow time is fixed at $t=0.01$, 
and the molecular dynamics parameters are set to 
$\Delta s = 0.02$ and $N_{\rm step}=50$. 
Computations are performed 
with a fixed number of threads $(=8)$, 
and the flow equations \eqref{flow_c}--\eqref{flow_ctn} 
are solved with DOPRI5(4) \cite{Hairer:2008}.

\begin{figure}[tb]
  \centering
  \includegraphics[width=75mm]{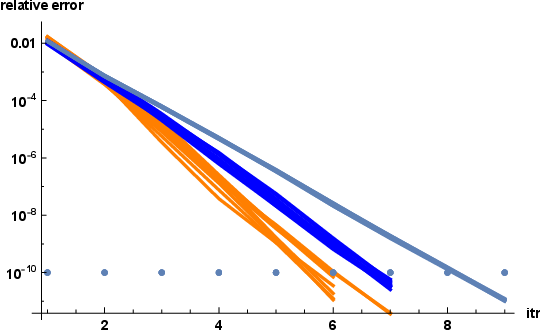}
  \caption{
    History of relative errors in BiCGStab 
    to solve the linear problem $A w_0=w$. 
    The data points at $L^2=16^2$ are joined by an orange line 
    for each $w\in T_z\bbC^N$ randomly generated (with a fixed $z$), 
    those at $L^2=128^2$ by a blue line, 
    and those at $L^2=512^2$ by a gray line. 
    We set the tolerance at $10^{-10}$. 
  }
  \label{fig:gt_bicgstabEF_convergence}
\end{figure}

\begin{figure}[tb]
  \centering
  \includegraphics[width=75mm]{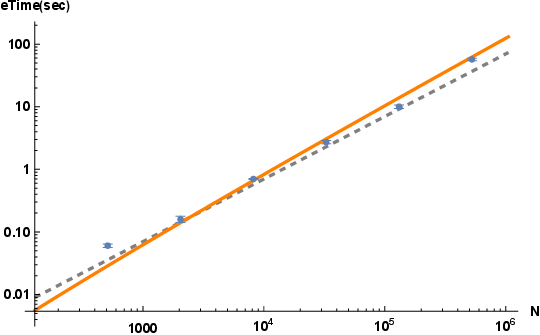}
  \caption{
    Elapsed time of the inversion of the linear problem 
    $A w_0 = w$ with BiCGStab 
    for lattices $L^2$ with $L=16,\,32,\,64,\,128,\,256,\,512$ 
    (note that $N=2 L^2$).
    The dashed gray line stands for $7\times 10^{-5} \times N$, 
    and the solid orange line for $9\times 10^{-6}\times N \log N$. 
  }
  \label{fig:gt_bicgstabEF_eTime}
\end{figure}

As discussed above, 
every iteration method utilizes the projection of a vector $w\in T_z\bbC^N$ 
onto the tangent space $T_z\Sigma$ 
and the normal space $N_z\Sigma$ [Eq.~\eqref{gt_projector}]. 
The dominant part in the computation 
is the inversion of the linear problem $A w_0 = w$ 
to find $w_0\in T_x\bbC^N$ 
for a given $w\in T_z\bbC^N$, 
and we use the BiCGStab method to solve this equation. 
Figure~\ref{fig:gt_bicgstabEF_convergence} shows 
the history of relative errors in BiCGStab, 
from which we find that 
the system size dependence of the convergence is very weak.
Figure~\ref{fig:gt_bicgstabEF_eTime} is the elapsed time 
to solve the linear equation. 
The iteration is terminated 
when the relative error falls below a prescribed tolerance $(= 10^{-10})$. 
The statistical errors are estimated 
from ten $w$'s randomly generated in $T_z\bbC^N$ with a fixed $z$. 
From the figure, 
the computational cost is expected to be in the range $O(N)\sim O(N \log N)$.

\begin{figure}[tb]
  \centering
  \includegraphics[width=75mm]{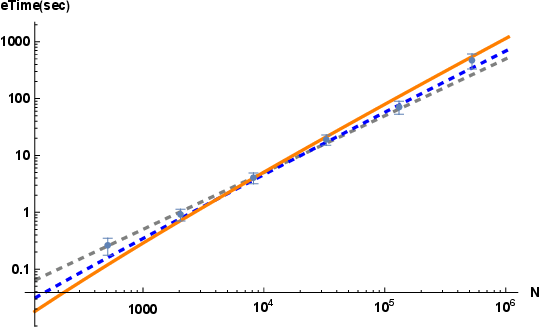}
  \caption{
    Elapsed time to solve Eq.~\eqref{gt_newton1} 
    with the simplified Newton method 
    using iterative solvers for orthogonal decompositions of vectors
    (Algorithm~\ref{alg:gt_rattle}). 
    The lattice size is $L^2$ with $L=16,\,32,\,64,\,128,\,256,\,512$ 
    (note that $N=2 L^2$).
    The dashed gray line stands for $5\times 10^{-4} \times N$, 
    the dashed blue line for $5\times 10^{-5}\times N \log N$, 
    and the solid orange line for $6\times 10^{-6}\times N (\log N)^2$. 
  }
  \label{fig:gt_solve_rattle_constraints_eTime}
\end{figure}

Figure~\ref{fig:gt_solve_rattle_constraints_eTime} is the elapsed time 
for a given $(z,\pi)\in T\Sigma$ 
to solve Eq.~\eqref{gt_newton1} with respect to $(u,\lambda)$ 
using the simplified Newton equation 
(Algorithm~\ref{alg:gt_rattle}) with iterative solvers for orthogonal decompositions of vectors.  
The physical and molecular dynamics parameters are the same as above. 
The iteration is terminated 
when $|B|$ $(B=z+\Delta z-\lambda-z_{\rm new})$ [Eq.~\eqref{gt_B}]
falls below a prescribed absolute tolerance $(= 10^{-10})$. 
The statistical errors are estimated 
from ten $(z,\pi)$'s randomly generated in $T\Sigma$. 
This shows that the computational cost 
would be in the range $O(N)\sim O(N\,(\log N)^2)$. 
The computational cost scaling is studied in more detail 
in a subsequent paper \cite{fn1}.
%

\subsection{Comparison between the simplified and the original RATTLE algorithms}
\label{sec:full_newton}

In this subsection, 
we compare the convergence of two algorithms 
for solving Eq.~\eqref{gt_newton1}. 
One is the simplified RATTLE algorithm, 
where the simplified Newton method (fixed-point method) 
is used [Eq.~\eqref{gt_newton3}] 
along with iterative solvers for orthogonal decompositions of vectors 
(Algorithm~\ref{alg:gt_decomposition} with iterative solvers). 
The other uses the standard Newton method (as in Ref.~\cite{Fukuma:2019uot}) 
with the following equation [Eq.~\eqref{gt_newton2}]:
\begin{align}
  &E_{\rm new} \Delta u + \Delta \lambda = B
  \nonumber
  \\
  &
  \left(
  \begin{array}{l}
    B =~ z + \Delta z - \lambda - z_{\rm new}\in \bbC^N, \\
    z_{\rm new}=z_t(x+u), \quad E_{\rm new} = (\partial z_t/\partial x)(x+u)
  \end{array}
  \right).
  \label{gt_newton2_app}
\end{align}
Introducing $\Delta\lambda_0\equiv F^{-1}\Delta \lambda$, 
we solve Eq.~\eqref{gt_newton2_app} 
with respect to $X=(\Delta u, \Delta \lambda_0)$ iteratively 
($\Delta\lambda$ is then obtained as $\Delta\lambda=F \Delta\lambda_0$). 
To do this, 
we first define an $\bbR$-linear map 
$\tilde{A}:\,T_{x+u}\Sigma_0 \oplus N_x\Sigma_0 \ni w_0
\mapsto
w\in T_{z_{\rm new}}\Sigma \oplus T_z\Sigma
$
by
\begin{align}
  w = \tilde{A}w_0 = E_{\rm new}v'_0 \oplus F n_0 \mbox{~~for~~}
  w_0 = v'_0 \oplus n_0,
\end{align}
where $E_{\rm new}v'_0 \in T_{z_{\rm new}}\Sigma$ 
and $F n_0 \in N_z\Sigma$ 
are obtained 
from $v'_0\in T_{x+u}\Sigma_0$ and $n_0\in N_x\Sigma_0$, 
respectively, 
by integrating two flow equations
from $x+u$ to $z_{\rm new}=z_t(x+u)$ 
and from $x$ to $z=z_t(x)$.
Now Eq.~\eqref{gt_newton2_app} takes the form $\tilde{A} X=B$ 
and can be solved iteratively by using the BiCGStab method 
as
\begin{align}
  \Delta u = {\rm Re\,}X,\quad
  \Delta \lambda_0 = i\,{\rm Im\,}X 
  \quad
  [\mbox{and thus~} \Delta \lambda = F\,(i\,{\rm Im\,}X) \,(= i\,E\,{\rm Im\,}X)].
\end{align}

\begin{figure}[tb]
  \centering
  \includegraphics[width=90mm]{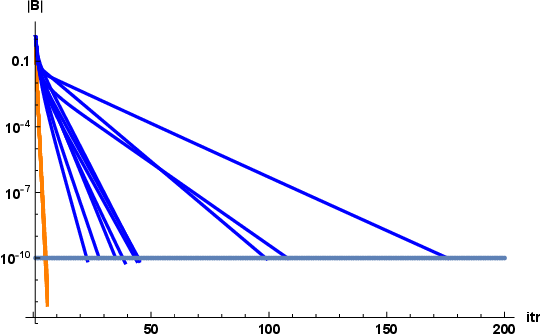}
  \caption{History of absolute errors $|B|$ 
    in solving Eq.~\eqref{gt_newton1} with two Newton methods iteratively 
    for ten different configurations $(z,\pi)\in T\Sigma$. 
    The lattice size is $32\times 32$. 
    For each configuration, 
    the data points obtained with the simplified Newton iteration
    [Eq.~\eqref{gt_newton3}]
    are joined by an orange line 
    while those with the standard Newton iteration 
    [Eq.~\eqref{gt_newton2} or Eq.~\eqref{gt_newton2_app}] by a blue line.}
  \label{fig:gt_rattle_simplified-full_convergence}
\end{figure}

\begin{figure}[tb]
  \centering
  \includegraphics[width=90mm]{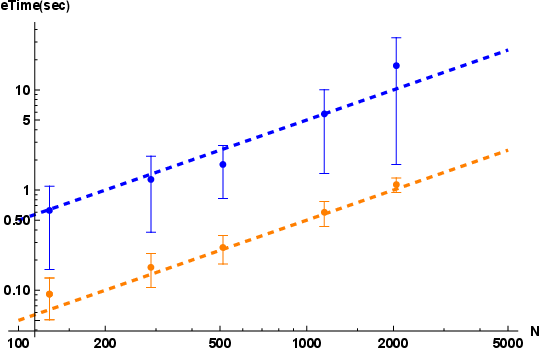}
  \caption{
    Elapsed times to solve Eq.~\eqref{gt_newton1} 
    with the simplified Newton iteration (orange) [Eq.~\eqref{gt_newton3}] 
    and the original Newton iteration (blue) [Eq.~\eqref{gt_newton2_app}]. 
    The lattice size is $L^2$ with $L=8,\,12,\,16,\,24,\,32$ 
    (note that $N=2 L^2$), 
    and the computation is made with a fixed number of threads ($=4$). 
    The dashed orange line stands for $5\times 10^{-4} \times N$, 
    and the dashed blue line for $5\times 10^{-3}\times N$. 
  }
  \label{fig:gt_rattle_simplified-full_eTime}
\end{figure}

Figure~\ref{fig:gt_rattle_simplified-full_convergence} 
shows the convergence of the iteration for two algorithms 
to solve Eq.~\eqref{gt_newton1}
on the lattice of size $32\times 32$ 
for ten different configurations $(z,\pi)\in T\Sigma$
used in Fig.~\ref{fig:gt_solve_rattle_constraints_eTime}. 
We see that the iteration of the simplified method (orange lines)
rapidly converges almost independently of configurations, 
while the original method (blue lines) 
requires many iterations to converge 
and the convergence varies in very different ways depending on configurations. 
Figure~\ref{fig:gt_rattle_simplified-full_eTime} 
shows the elapsed times for two algorithms  
on the lattice of size $L^2$ with $L=8,\,12,\,16,\,24,\,32$ 
for ten different configurations. 
We see that both exhibit the computational cost scaling $\sim O(N)$, 
but the simplified algorithm is much faster than the original algorithm. 
The rate of improvement has large deviations, 
reflecting the serious dependence of the original algorithm on configurations.
%

\section{Conclusion}
\label{sec:conclusion}

We have developed a simplified algorithm for the GT-HMC and the WV-HMC methods, 
by adopting a simplified Newton method 
(worldvolume version of the fixed-point method of Ref.~\cite{Fujii:2013sra})
in determining the Lagrange multiplies of RATTLE 
with iterative solvers for orthogonal decompositions of vectors. 
Using as a benchmark model 
the complex scalar field theory at finite density,%
\footnote{ 
  A detailed investigation of the model with the simplified GT/WV-HMC algorithm 
  is carried out in Ref.~\cite{fn1}.
} 
we performed a numerical test 
for the convergence of the simplified RATTLE algorithm. 
We found that 
the convergence depends on the system size only weakly 
and the computational cost is nearly $O(N)$ 
at each molecular dynamics step. 
 
In subsequent papers \cite{fn1,fn2}, 
we apply the current algorithm to various quantum field theories. 
The target models of the WV-HMC method 
can be classified into two categories. 
The first consists of models whose action is purely local, 
for which the Hessian $(H_{ij}=\partial_i\partial_j S)$
appearing in flow equations are sparse matrices. 
A typical example is the complex scalar field theory at finite density, 
and will be studied in Ref.~\cite{fn1}. 
The other treats models whose bosonized actions include nonlocal terms 
such as the logarithm of the fermion determinant. 
A study of dynamical fermion systems with the WV-HMC method 
will be made in Ref.~\cite{fn2}. 

Models discussed above have configuration spaces of flat geometry. 
Actually, the WV-HMC method can also be generalized to models 
whose configuration spaces are group manifolds. 
This will be discussed in Ref.~\cite{mf}. 

Besides applying the WV-HMC method to various models, 
we believe that it is also important to keep improving the algorithm itself. 
It should be interesting to combine the WV-HMC algorithm 
with other methods towards solving the sign problem, 
such as the complex Langevin method and/or the tensor network method. 
It is also interesting to incorporate machine learning techniques 
in order to further reduce the computational cost. 

One of the most important projects in the near future 
is to develop the Monte Carlo algorithm 
to study the real-time dynamics of quantum many-body systems 
(see, e.g., Refs.~\cite{Alexandru:2016gsd,Mou:2019tck,Mou:2019gyl,Nishimura:2023dky} 
for attempts based on the generalized thimble method). 
A study based on the WV-HMC is now in progress 
and will be reported elsewhere.

\section*{Acknowledgments}

The author thanks 
Sinya Aoki, Ken-Ichi Ishikawa, Issaku Kanamori, Yoshio Kikukawa, 
Nobuyuki Matsumoto and Naoya Umeda 
for valuable discussions, 
and especially Yusuke Namekawa 
for collaboration and comments on the manuscript. 
This work was partially supported by JSPS KAKENHI 
(Grant Numbers 20H01900, 23H00112, 23H04506) 
and by MEXT as 
``Program for Promoting Researches on the Supercomputer Fugaku'' 
(Simulation for basic science: approaching the new quantum era, JPMXP1020230411).


\appendix

\section{Proof of Eq.~\eqref{wv_force}}
\label{sec:wv_force}

We start from the expression
\begin{align}
  \overline{\partial V} 
  = \frac{1}{2}\,\overline{\partial S}
  + W'(t)\,\overline{\partial t}
  = \frac{1}{2}\,\xi + W'(t)\,\overline{\partial t}.
\end{align}
We decompose $\overline{\partial t} \in T_z\bbC^N$ 
at $z\in \Sigma_t \,(\subset \calR)$
into the form
\begin{align}
  \overline{\partial t} 
  = (\overline{\partial t})_\parallel + (\overline{\partial t})_\perp
  = v + c\,\xi_n + (\overline{\partial t})_\perp
\end{align}
with $v=v^a E_a\in T_z\Sigma_t\,(\subset T_z\calR)$. 
Note that $\xi_n\in N_z\Sigma_t\cap T_z\calR$ and 
$(\overline{\partial t})_\perp\in N_z\calR$. 
In the following, 
we will show that (1)~$v=0$ and (2)~$c=1/(2\,\langle \xi_n,\xi_n \rangle)$. 
This completes the proof 
because $(\overline{\partial t})_\perp\in N_z\calR$ can be absorbed 
into $\lambda$ in Eq.~\eqref{wv_rattle1}.

\underline{(1)} 
For $\forall z\in\Sigma_t$, $\forall u=u^a E_a\in T_z\Sigma_t$
and an infinitesimally small $\epsilon$, 
we have 
\begin{align}
  t(z,\bar z) = t(z+\epsilon u, \bar z+\epsilon \bar u)
  = t(z,\bar z)
    + \epsilon\,[u^i\,\partial_i t
                 + \overline{u^i}\,\overline{\partial_i t}]
    + O(\epsilon^2),
\end{align}
and thus,
\begin{align}
  0 = u^i\,\partial_i t + \overline{u^i}\,\overline{\partial_i t} 
    = 2\,\langle u,\overline{\partial t} \rangle
    = 2\,\langle u,v \rangle
    = 2\,u^a \gamma_{ab} v^b.
\end{align}
This means that $v=0$ due to the nondegeneracy of $\gamma_{ab}$. 

\underline{(2)} 
By using the orthogonality, 
$c$ is given by 
\begin{align}
  c = \frac{\langle \xi_n, \overline{\partial t} \rangle}
      {\langle \xi_n,\xi_n \rangle}.
\end{align}
Here, noting that
\begin{align}
  1 = [t(z,\bar z)]^\centerdot
    = \dot{z}^i\,\partial_i t 
      + \overline{\dot{z}^i}\,\overline{\partial_i t}
    = \xi^i\,\partial_i t + \overline{\xi^i}\,\overline{\partial_i t}
    = 2\,\langle \xi,\overline{\partial t} \rangle
    = 2\,\langle \xi_n,\overline{\partial t} \rangle
  \quad(\because v=0),
\end{align}
we have $\langle \xi_n,\overline{\partial t} \rangle = 1/2$.
We thus obtain $c = 1/(2\,\langle \xi_n,\xi_n \rangle )$.

\baselineskip=0.9\normalbaselineskip



\end{document}